\documentclass[english,amsmath,amssymb,aps,prl,showpacs,twocolumn]{revtex4-1}
\pdfoutput=1
\usepackage[T1]{fontenc}
\usepackage{amsmath}
\usepackage{amssymb}
\usepackage{graphicx}
\usepackage{esint}
\usepackage{caption}
\usepackage{ulem}
\usepackage{subcaption}
\usepackage{color}
\newcommand{\be}{\begin{equation}}
\newcommand{\ee}{\end{equation}}
\newcommand{\medio}[1]{\left\langle #1 \right\rangle}

\makeatletter

\usepackage{babel}

\makeatother

\begin{document}

\title{Cloaking Resonant Scatterers and Tuning Electron Flow in Graphene}
\author{Diego Oliver$^1$, Jose H. Garcia$^1$, Tatiana G. Rappoport$^1$, N. M. R. Peres$^2$, and Felipe A. Pinheiro$^1$}
\affiliation{$^1$Instituto de F\'isica, Universidade Federal do Rio de Janeiro, Caixa Postal 68528, 21941-972 Rio de Janeiro RJ, Brazil}
\affiliation{$^2$Department of Physics and Center of Physics, University of Minho, P-4710-057 Braga, Portugal}

\begin{abstract} 
We consider resonant scatterers with large scattering cross-sections 
in graphene that are produced by a gated disk or a vacancy, and show 
that a gated ring can be engineered to produce an efficient 
electron cloak. We also demonstrate that this same scheme can be 
applied to tune the direction of electron flow. Our analysis 
is based on a partial-wave expansion of the electronic wave-functions 
in the continuum approximation, described by the Dirac equation. 
Using a symmetrized version of the massless Dirac equation, we derive a general
condition for the cloaking of a scatterer by a potential with radial symmetry.
We also perform tight-binding calculations to show that our findings 
are robust against the presence of disorder in the gate potential.

\end{abstract}

\maketitle

\section{Introduction}

Analogies between light and electrons are known to be at
the origin of a plethora of interesting physical phenomena in 
condensed matter and in optics. Interference effects such as 
Anderson localization~\cite{wiersma2013} and
 Fano resonances~\cite{fano2010} are just two 
examples of effects that have been observed 
in both areas of Physics. In the last decade the advent of metamaterials 
and the isolation graphene have been fostering the discovery of novel fundamental phenomena 
and applications based on the analogy between light and electrons.

On one hand, the low-energy electronic excitations of graphene~\cite{novoselov2004}, 
are described by Dirac fermions with linear 
dispersion relation similar to photons~\cite{novoselov2005}. 
This ``optics-like" dispersion relation has been
exploited in the development of electronic analogues of photonic 
devices such as beam-splitters, wave-guides, and Fabry-Perot 
interferometers in ballistic graphene~\cite{zhang2009,rickhaus2013}. 
Graphene-based electronic analogues of optical applications grounded on 
metamaterials, such as perfect lenses, have been also proposed~\cite{cheianov2007,silveirinha2013}. 

On the other hand progresses in the field of metamaterials~\cite{zheludev2012} 
have allowed the development of electromagnetic cloaks to achieve invisibility, 
and stimulated the quest for their electronic analogues. Among the existing approaches
 to achieve invisibility one can highlight the coordinate-transformation 
method~\cite{pendry2006,leonhardt2006,schurig2006} and the scattering cancellation 
technique~\cite{alu2005,alu2008,edwards2009,filonov2012,chen2012,rainwater2012,KortKamp2013,nicorovici1994}. 
The coordinate-transformation method requires metamaterials with anisotropic and inhomogeneous profiles, which are able to bend the 
incoming electromagnetic radiation around a given region of space, rendering it invisible to an external observer. 
This method has been first realized experimentally for microwaves~\cite{schurig2006}, and later for
 infrared and visible frequencies~\cite{valentine2009,ergin2010}. Extensions of this method to
 matter waves do exist~\cite{zhang2008,greenleaf2008,lin2010,lin2011}, proving that it is possible 
to reroute the probability flow around a scatterer. 
However, as in the electromagnetic case these electronic analogues require cloaking layers 
with extreme material parameters, such as anisotropic potential profiles and/or infinite
 values for the effective mass, imposing limitations to practical implementations in 
condensed matter systems. Alternatively, the scattering cancellation technique for 
electromagnetic waves relaxes many of these constraints in terms of metamaterial requirements.
 It uses an homogeneous and isotropic plasmonic layer in which the induced polarization vector
 is out-of-phase with respect to the electric field so that the in-phase contribution given by 
the cloaked object may be partially or totally canceled~\cite{alu2005,alu2008,chen2012,wilton2013}.
 The scattering cancelation technique was first experimentally implemented for
 microwaves~\cite{edwards2009,rainwater2012}, and later extended to acoustic~\cite{guild2011}, and matter
 waves~\cite{liao12,fleury2012,fleury2013}. One of the proposals for matter waves cloak, based on 
the expansion of partial waves and later extended to graphene~\cite{liao13}, 
demonstrates that core-shell nanoparticles embedded in a host semiconductor with a size 
comparable to electronic wavelengths can be ``invisible'' {to incoming electrons}~\cite{liao12}. 
To achieve this effect the contribution of the first two partial waves must vanish simultaneously 
so that the total scattering cross section can be smaller than 0.01\% of the physical cross section. 
As we shall see, a similar effect can be exploited in graphene for the cloaking of
electronic  scatterers.   

Charge carriers in graphene have high mobility, and their peculiar 
properties cause small transport cross-sections as a result of 
back-scattering suppression. However, different types of defects, 
such as vacancies and covalently bonded adatoms, can form resonant states at, or close to,
 the Dirac point~\cite{pereira2006,peres2010,wehling2010}. Resonant states increase the scattering 
cross-section by many orders of magnitude and hence may limit the carrier's mobilities. 
They can also can be used to modify the electronic properties in the vicinity of
 the Dirac point. 

In the present article we combine the idea of ``electronic cloaking'' with 
graphene's peculiar electronic properties to propose an alternative and 
efficient scheme for achieving electronic invisibility of resonant scatterers. 
{We show that sharp transitions between invisible and resonant states occur for 
some values of the potential, which can be viewed as an electronic analogue of 
optical comblike states in coated scatterers~\cite{monticone2013}}. We consider 
two different resonant scatterers: an engineered resonant state produced by a top 
gate and a vacancy.  We demonstrate that the ``electron cloak'' scheme 
proposed in Refs.~\cite{liao12,liao13} is strongly facilitated by the fact that the first two phase shifts, which give the dominant contribution in the treatment of electron scattering by a core-shell impurity, are identical for graphene. As a result, one just need to tune one of the scattering potentials to vanish the scattering cross-section, in contrast to the case where the impurity is embedded in a host semiconductor~\cite{liao12}.
When the sub-lattice symmetry is broken in graphene, {\it e.g.} as in the case a vacancy, 
the first two phase shifts are no longer equal so that energy selective cloaking 
is more difficult to achieve. However, it is highly efficient at the Dirac point, 
where the resonant state is located. We also derive a general
condition for the cloaking of a scatterer by a potential with radial symmetry using a symmetrized version of the massless Dirac equation.
We extend the analysis to tight-binding calculations and
show that our results are robust against the presence of disorder.  
Our findings reveal that electron cloaks are not only more easily 
implemented in graphene, but also that they could be applied to 
selectively tune the direction of electron flow in this material.  
We suggest that 
new devices may be engineered by exploiting the cloaking effect.

The article is organized as follows: 
in section \ref{potential} we introduce the 
partial-waves formalism for graphene, and find  
the conditions for resonant scattering produced by 
a gate potential and a vacancy.  In section \ref{cloaking} 
we analyze the effect of surrounding the two types of resonant 
scatterers with a potential ring, leading to cloaking. 
We also derive a general condition for cloaking by a potential
of radial symmetry; the example of cloaking of a void by a $\delta-$function
potential is provided. In section \ref{TB} we perform tight-binding 
calculations  to demonstrate that our findings are 
robust against disorder in the ring, even in the presence of non-perfect potentials. 

\section{ Single potential and the resonant scattering regime \label{potential}}

We consider a monolayer of graphene with a gated disk that works as a scattering center, and we choose its potential to maximize its total cross section. This can be achieved in the resonant scattering regime.
Our starting point is the continuum-limit Hamiltonian of graphene
\be {\cal H}_{0}=\hbar v_{F}(\tau_{z}\sigma_{x}p_{x}+\sigma_{y}p_{y}), \ee
where $\mathbf{p}=(p_{x},p_{y})$ is the momentum operator
around one of the two inequivalent Dirac points $K$ and $K^{\prime}$,
$v_{F}\approx10^{6}$~m/s is the Fermi velocity, $\boldsymbol{\sigma}$
and $\boldsymbol{\tau}$ denote Pauli matrices, with $\sigma_{z}=\pm1$
($\tau_{z}=\pm1$) describing states on A-B sublattice (at $K$-$K^{\prime}$).
For such large scatterers intervalley scattering is negligible and, in the long wavelength limit, for potentials with radial symmetry, the scatterer is described by ${\cal H_V}=V\Theta(R-r)$ where $V$ is a constant, $R$ is the radius of the scatterer and $\Theta(.)$ is the Heaviside function.
Here we use the partial-wave method to  calculate the cross section of scattered waves by a radially symmetric potential. 
In what follows, we derive the partial-wave scattering amplitudes and from their knowledge,  all gauge-invariant quantities can be determined unequivocally.
In cylindrical coordinates, the components of the graphene spinor $\Psi(\mathbf{r})=(\psi_{A}({\mathbf{r}}),\psi_{B}({\mathbf{r}}))^{T}$
are decomposed in terms of radial harmonics
\begin{align}
 & \psi_{A}({\mathbf{r}})=\sum_{m=-\infty}^{\infty}g_{m}^{A}(r)e^{im\theta}\\ 
 &\psi_{B}^{}({\mathbf{r}})=\sum_{m=-\infty}^{\infty}g_{m}^{B}(r)e^{i(m+1)\theta}\,,\label{eq:Amp_A}
\end{align}
where $\theta\equiv\textrm{arg}(k_{x}+ik_{y})$, $\mathbf{k}$ is
the wavevector, $m$ is the angular momentum quantum number and
 $A(B)$ represents the sublattice. After separating the variables of the graphene
plus impurity Hamiltonian ${\cal H_0}+{\cal H_V}$, we obtain two coupled first order
equations for the radial functions $g_{m}^{A}(r)$ and $g_{m}^{B}(r)$, where $\boldsymbol{\mathbf{\sigma}},\mathbf{\boldsymbol{\tau}}$
are Pauli matrices for sublattice and valley
respectively. $\tau_{z}$ is conserved and thus we can focus on states
at $K$ valley; scattering amplitudes for states
at $K^{\prime}$ are quantitatively the same. In a partial-waves expansion,
the asymptotic form for the spinor wave-function is given by~\cite{novikov2007,peres2010}
\begin{align}
\psi_{\lambda,\mathbf{k}}({\bf r}) & =\left(\begin{array}{c}
1\\
\lambda
\end{array}\right)e^{ikr\cos(\theta)}+\frac{f(\theta)}{\sqrt{-ir}}\left(\begin{array}{c}
1\\
\lambda e^{i\theta_{{\bf k}}}
\end{array}\right)e^{ikr}\notag\nonumber \\
 & \label{eq:psiKas}
\end{align}
where $\lambda=\pm1$ denotes the carrier polarity and $f(\theta)$ is scattering amplitude. Using a partial wave analysis, we can relate $f(\theta)$ with the phase-shifts $\delta_m$.
\begin{equation}
f(\theta)=\sqrt{\frac{2}{\pi k}}\sum_{m=-\infty}^{\infty}e^{im\theta}e^{i\delta_m}\sin{\delta_m}.
\label{eq.4}
\end{equation}

The differential cross section is given by $\frac{d\sigma}{d\theta}=|f(\theta)|^2$ and the total cross section $\sigma(kR)$  and transport cross-section $\sigma_T(kR)$ are given by 
\begin{align}
&\sigma(kR)=\int \frac{d\sigma}{d\theta} d\theta=\frac{4}{k}\sum_{m=-\infty}^{\infty}\sin^2{\delta_m}, \\
&\sigma_T(kR)=\int \frac{d\sigma}{d\theta} [1-\cos(\theta)]d\theta=\frac{2}{k}\sum_{m=-\infty}^{\infty}\sin^2{(\delta_m-\delta_{m+1})}.
\label{eq.5}
\end{align}
 The d.\,c.
 conductivity of graphene layer at zero temperature can be obtained directly from  $\sigma_T(kR)$~\cite{peres2010}.

To identify the phase-shifts, we write the spinors for the region inside and outside the potential as a superposition of angular harmonics. For $r>R$, we have $E=\lambda \hbar v_{F}k_{out}$ and the partial-wave $m$
is a sum of an incoming and a scattered wave according to 
\begin{align}
\psi_{m}^{\text{out}}(r,\theta) & =A_{m}\left(\begin{array}{c}
J_{m}(k_{\text{out}}r)e^{im\theta}\\
i\lambda_{\text{out}}J_{m+1}(k_{\text{out}}r)e^{i(m+1)\theta}
\end{array}\right)\nonumber\\&+B_{m}\left(\begin{array}{c}
Y_{m}(k_{\text{out}}r)e^{im\theta}\\
i\lambda_{\text{out}} Y_{m+1}(k_{\text{out}}r)e^{i(m+1)\theta}
\end{array}\right),
\label{eq.6}
\end{align}
whereas for $r<R$ we have 
\begin{equation}
\psi_{m}^{\text{in}}(r,\theta)=C_{m}\left(\begin{array}{c}
J_{m}(k_{\text{in}} r)e^{im\theta}\\
i\lambda_{in} J_{m+1}(k_{\text{in}} r)e^{i(m+1)\theta}
\end{array}\right),
\label{eq.7}
\end{equation}
where $k_{\text{in}}\equiv |E-V|/\hbar v_{F}$. The boundary
condition $\psi_{m}^{\text{in}}(R,\theta)=\psi_{m}^{\text{out}}(R,\theta)$ leads to two equations. The solution of these equations determine the ratio $B_{m}/A_{m}$.  It is straightforward to show that the phase-shift $\delta_{m}$ for partial-wave $m$ relates to the $B_{m}/A_{m}$ according to $B_{m}/A_{m}=-\tan(\delta_m)$ (for more details about partial-waves in graphene, see references~\cite{novikov2007, peres2010,Ferreira11_sup}).

The Dirac equation with a radial scalar potential is symmetric under exchanging
 $g_{m}^{A}$ and $g_{-m-1}^{B}$, which leads to the absence of back-scattering.
The latter also corresponds to the relation $\delta_{-m}=\delta_{m+1}$ between phase-shifts.
For small energies $kR\ll1$, channels $m=-1,0$ give the main
contributions to $f(\theta)$.

\begin{figure}[h]
\includegraphics[width=0.9 \columnwidth]{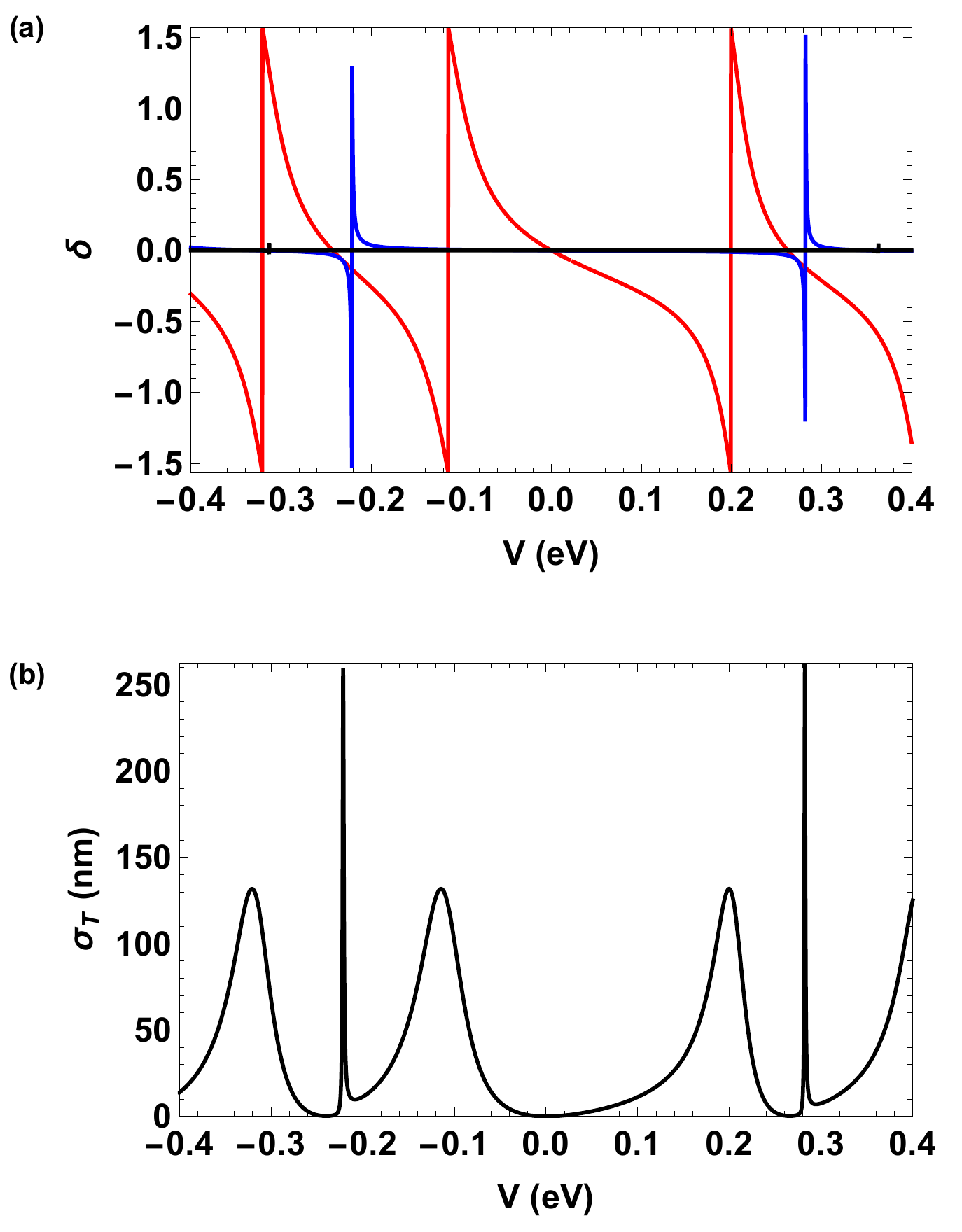}

\caption{\label{fig.1} (a) phase shifts  $\delta_0$ (black), $\delta_1$ (red) and $\delta_2$ (blue) and (b) transport cross-section $\sigma_T $ ( 6 first partial waves are considered) as a function of the electrostatic potential $V$ for $E=0.02$ eV and $R=10$ nm.}
\end{figure}

For a fixed energy of the incoming electron, we can tune the potential $V$ to 
produce a resonant scatterer that will effectively trap the electrons inside the disk, which occurs for phase-shifts of $\pm \pi/2$. Figure \ref{fig.1} (a) shows the 
phase shifts of three partial waves; there are two resonances at different values of electrostatic potential, associated to the contributions of $\delta_0=\delta_{-1}$  and $\delta_1=\delta_{-2}$.  

The transport cross-section as a function of the potential (Figure \ref{fig.1} (b)) presents pronounced peaks {that} occur exactly at the maximum of the phase- shifts, characterizing the resonant scattering regime. {Figure \ref{fig.1}(b) reveals that the transport cross-section exhibits asymmetric, Fano-like resonances associated to the modes for $m=-2,1$.} 

\subsection{\textmd Circular void  : another source of resonant scattering}
\begin{figure}[h]
\centering
\includegraphics[scale=0.5]{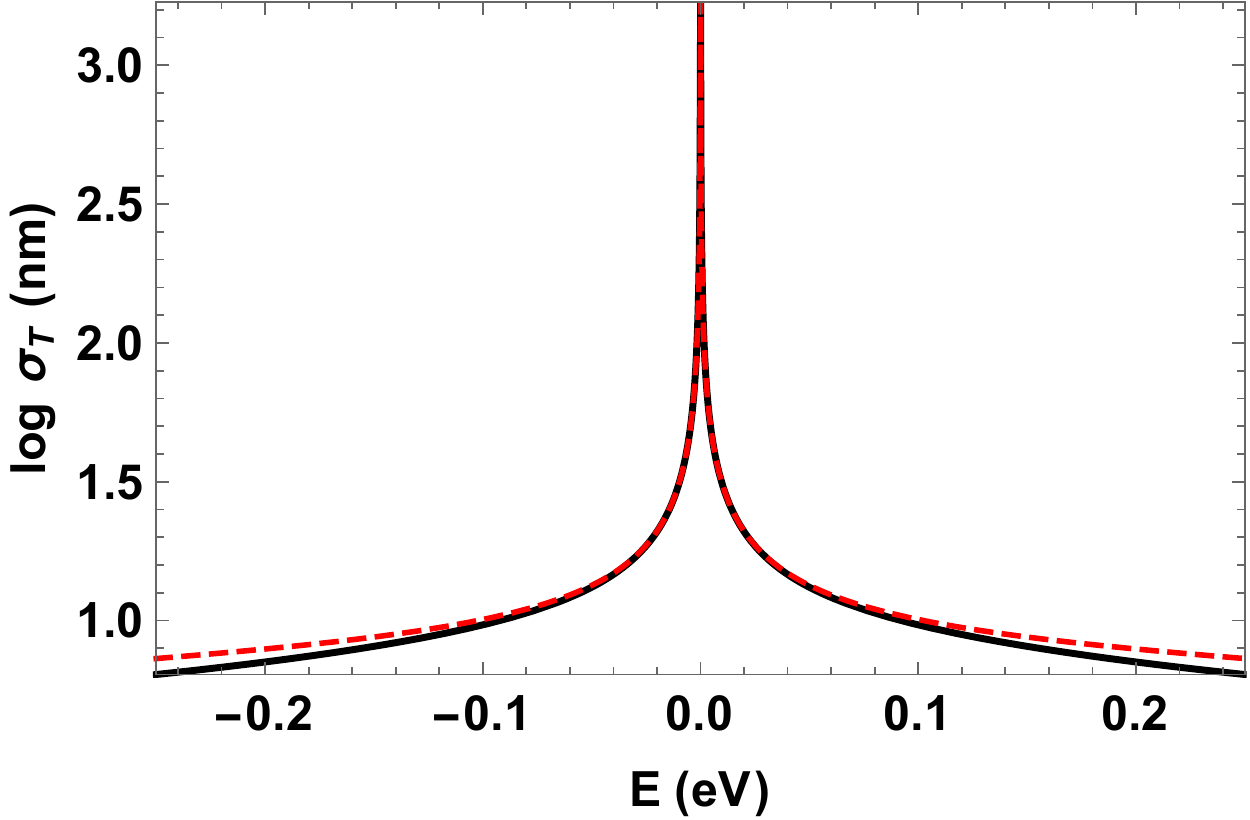}
\caption{Transport cross-section for a vacancy of radius $R_1=1 \;\text{nm}$ as a function $E$ considering the 6 first partial waves  (solid) and and a single partial-wave $\delta_0$ (dashed) .}
\label{fig.2}
\end{figure}

There are various ways to induce resonant scattering in graphene. One possibility is the creation of vacancies that give rise to zero-mode states~\cite{pereira2006,mayou2013,cresti2013}.  
Vacancies can be engineered  by irradiating high-energy ions~\cite{irrad1,irrad2}. This method allows the controlled incorporation of 
resonant scatterers, and can be used to taylor the electrical, mechanical, and magnetic properties of graphene~\cite{irrad1,irrad2}. 
A vacancy in graphene can be modeled by a void with 
zig-zag edges~\cite{pacopartial,Ferreira11_sup}. 
The boundary conditions for the wave-function at a 
void of radius $R$ with zig-zag edges is given by $\psi_{A}(r=R)=0$ 
for the $A$  spinor component. In this case, if we apply the boundary 
conditions to Eq.~\ref{eq.6}, we obtain
\begin{equation}
\delta_m=-\arctan \left(\frac{Y_m({k}_{\text{out}}R)}{J_m({k}_{\text{out}}R)} \right)
\label{eq.12}
\end{equation} 
where $E={\lambda}_{\text{out}} \hbar v_F {k}_{\text{out}}$. The zig-zag edges break sub-lattice symmetry 
and the relation between the $\delta_m$ and $\delta_{-m-1}$ is not valid anymore. 
As it can be seen from Fig.~\ref{fig.2}, the main contribution to the scattering 
cross-section at low energies is given by $\delta_0$~\cite{Ferreira11_sup}, 
that gives rise to resonant scattering for $E\rightarrow 0$.  

\section{\textmd{Cloaking resonant scatterers}\label{cloaking}}
\begin{figure}[h]
\centering
\includegraphics[clip,width=0.85\columnwidth]{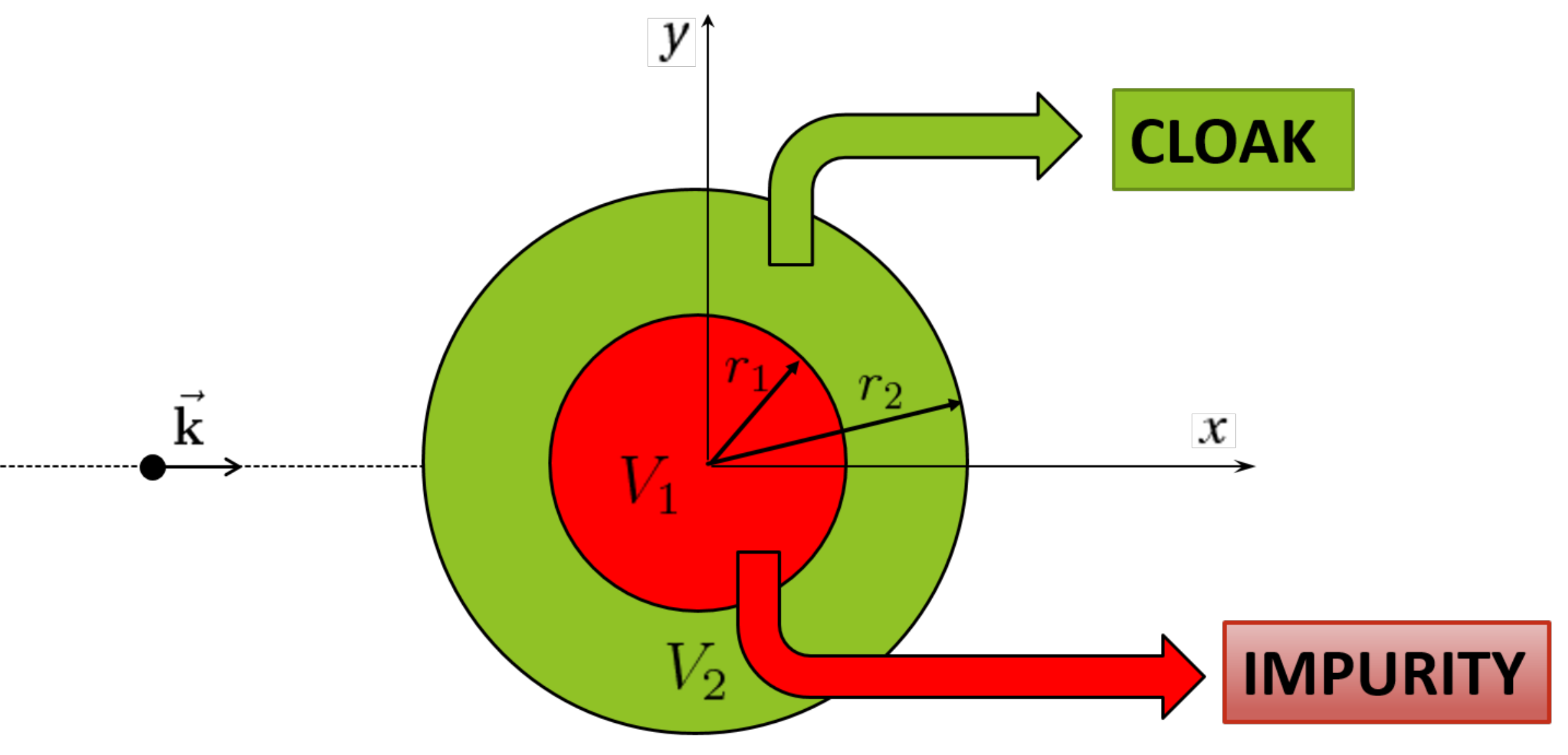}
\caption{Sketch of the problem, showing the incoming electron, the impurity, and the cloak.}
\label{fig.3}
\end{figure}

After preparing the system in the resonant scattering regime, 
which maximize the transport cross-section, we now proceed to discuss the
 cloaking scheme for the scalar potential. We keep the original potential 
$V$ (now called $V_1$) in the disk of radius $R$ (now called $R_1$) , and
 include a new potential $V_2$ in a ring of internal radius $R_1$, and 
external radius $R_2$ that is used to tune the cloaking mechanism
 (see Fig.~\ref{fig.3}). The basic idea is the following: we change  $V_2$ 
in such a way that $\delta_m$ vanishes, so that $\sigma\sim0$. Differently
 from a normal 2DEG, where it is necessary to tune $V_1$ and
$V_2$ to produce $\delta_0=\delta_{1}=0,\pm\pi$~\cite{liao12}, here 
{it suffices to modify} one of the phase-shifts, as $\delta_0=\delta_{-1}$ 
are the main contributors to the cross-section; hence when they are zero, 
the total cross-section goes effectively to zero (it is reduced by several orders of magnitude).

We perform the same type of calculations described in the previous section but 
instead of defining the wave-functions in two regions, we now have three:
Outside the potentials (region 3), for $r>R_2$, we have 
\begin{align}
\psi_{m}^{3}(r,\theta) & =A_{m}\left(\begin{array}{c}
J_{m}(k_{3}r)e^{im\theta}\\
i\lambda_{3} J_{m+1}(k_{3}r)e^{i(m+1)\theta}
\end{array}\right)  \nonumber \\ 
& +B_{m}\left(\begin{array}{c}
Y_{m}(k_{3}r)e^{im\theta}\\
i\lambda_{3} Y_{m+1}(k_{3}r)e^{i(m+1)\theta}
\end{array}\right).
\label{eq.9}
\end{align}

For $R_2<r<R_1$ (region 2) we have 
\begin{align}
\psi_{m}^{2}(r,\theta) & =C_m\left(\begin{array}{c}
J_{m}(k_{2}r)e^{im\theta}\\
i\lambda_{2} J_{m+1}(k_{2}r)e^{i(m+1)\theta}
\end{array}\right) \nonumber \\
&+D_{m}\left(\begin{array}{c}
Y_{m}(k_{2}r)e^{im\theta}\\
i\lambda_{2} Y_{m+1}(k_{2}r)e^{i(m+1)\theta}
\end{array}\right),
\label{eq.10}
\end{align}
whereas for $r<R_1$ (region 1) we have 
\begin{equation}
\psi_{m}^{1}(r,\theta)=E_{m}\left(\begin{array}{c}
J_{m}(k_{1} r)e^{im\theta}\\
i\lambda_{1} J_{m+1}(k_{1} r)e^{i(m+1)\theta}
\end{array}\right),
\label{eq.11}
\end{equation}
where $k_1\equiv |E-V_1|/ \hbar v_{F}$, $k_2\equiv |E-V_2|/\hbar v_{F}$ and $k_3\equiv |E|/ \hbar v_{F}$.
The boundary
conditions $\psi_{m}^{3}(R_2,\theta)=\psi_{m}^{2}(R_2,\theta)$ and $\psi_{m}^{1}(R_1,\theta)=\psi_{m}^{2}(R_1,\theta)$ lead to
four equations. The solution of these boundary conditions determine  $B_{m}/A_{m}$ and consequently $\delta_m$.

To highlight the effects of cloaking, we choose $V_1$ to maximize 
the initial cross-section, which is related to the resonant scattering 
through the disk of ratio $R_1$. Then, we tune the potential of the ring 
$V_2$ and its radius $R_2$ in order to minimize the cross-section. Let us
 start with the situation described in the Figure \ref{fig.1}, where we have
 a disk with radius $R_1=10\;\text{nm}$ and, for an incident beam with energy
 $E=0.02\;\text{eV}$ (a value close to the Dirac point, which can be easily
accessible for graphene on h-BN), resonant scattering occurs for some specific
 values of $V_1$. For $V_1=199\;\text{meV}$ and $V_1=-114\;\text{meV}$,  
the broad peaks in the transport cross-section are related to the resonance 
of the low-order phase-shift, $\delta_0=\delta_{-1}$. On the other hand, 
for $V_1=282\;\text{meV}$ and $V_1=221\;\text{me}$, we can see a sharp peak 
in consequence of the resonance of the phase-shifts $\delta_1=\delta_{-2}$. 
Next, we tune the potential and the radius of the ring to investigate 
the maximal decrease in $\sigma_T$ in these particular situations.  To identify
the best cloaking scenarios, we analyze the behavior of the cloaking 
efficiency, which is the ratio between the transport cross-sections with, $\sigma_T$, and without,
$\sigma_{T0}$,
 the ring, $\sigma_T/\sigma_{T0}$.

\begin{figure}[h]
\centering
\includegraphics[clip,width=0.95\columnwidth]{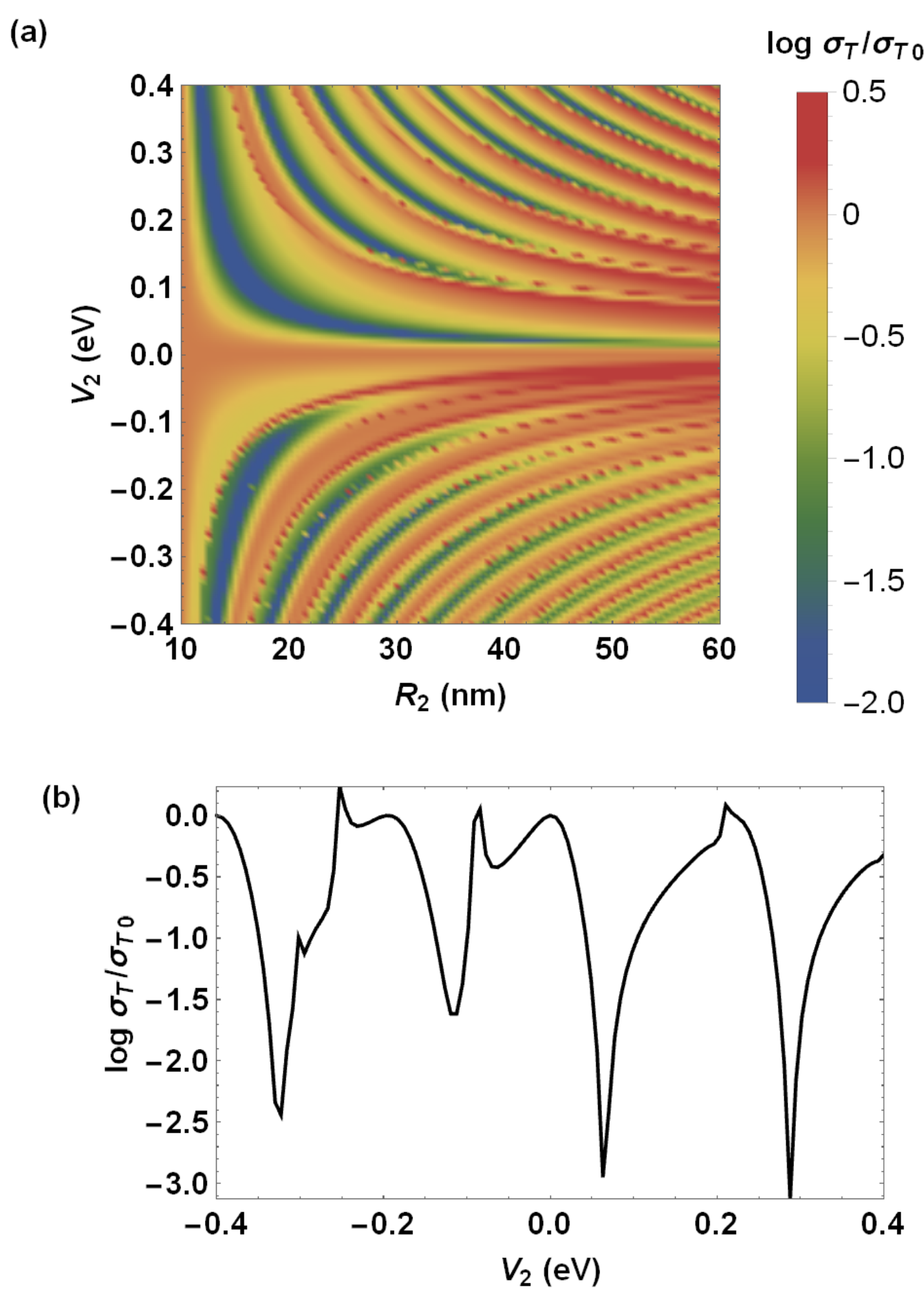}
\caption{(a) Cloaking efficiency as a function of $V_2$ and $R_2$ for  $E=0.02\;\text{eV}$, $R_1=10\;\text{nm}$ and $V_1=-114\;\text{meV}$. (b) Cloaking efficiency for the same parameters 
of (a) and $R_2=20$ nm.} 
\label{fig.4}
\end{figure}

Figure \ref{fig.4}(a) shows a density plot of the cloaking efficiency as a function of the potential $V_2$  and radius $R_2$ of the
 ring, for a fixed value of $E$ and of the transport cross-section for the resonant scatterer 
${\sigma_T}_{0}=132\;\text{nm}$ ($V_1=-114\;\text{meV}$). Panel  \ref{fig.4}(b), which is a lateral cut of  Figure \ref{fig.4} (a) for 
$R_2=20$ nm, reveals an impressive suppression of the transport cross-section of 3 orders of magnitude.

In view of these results, we conclude that the cloak is
very efficient in this setup.  Here, we consider a scenario where both the
resonant scattering and the cloaking are engineered with the 
use of tunable back and top gates~\cite{liao13}. In this particular case, it is necessary to work with disks and rings with radius in the order of tens of nm.  Best cloaking efficiency, of the order of $10^{-5}$, can be obtained with thin layers with a distance of $\sim 2$ nm between internal and external radius. {In spite of the fact that geometries with these sizes may be difficult to fabricate,} fortunately there are some regions in the parameters' space that provide a very effective cloaking for more realistic values of $R_2$ and $V_2$.

Moreover, for fixed $R_2$, there are large oscillations of the value of the
 transport cross-section as a function of the potential of the ring. These variations in the transport cross-section allow us to drastically change the transport regime: from cloaking to resonant scattering by applying small changes in the gate voltage $V_2$. 

\begin{figure}[h]
\centering
\includegraphics[clip,width=0.95\columnwidth]{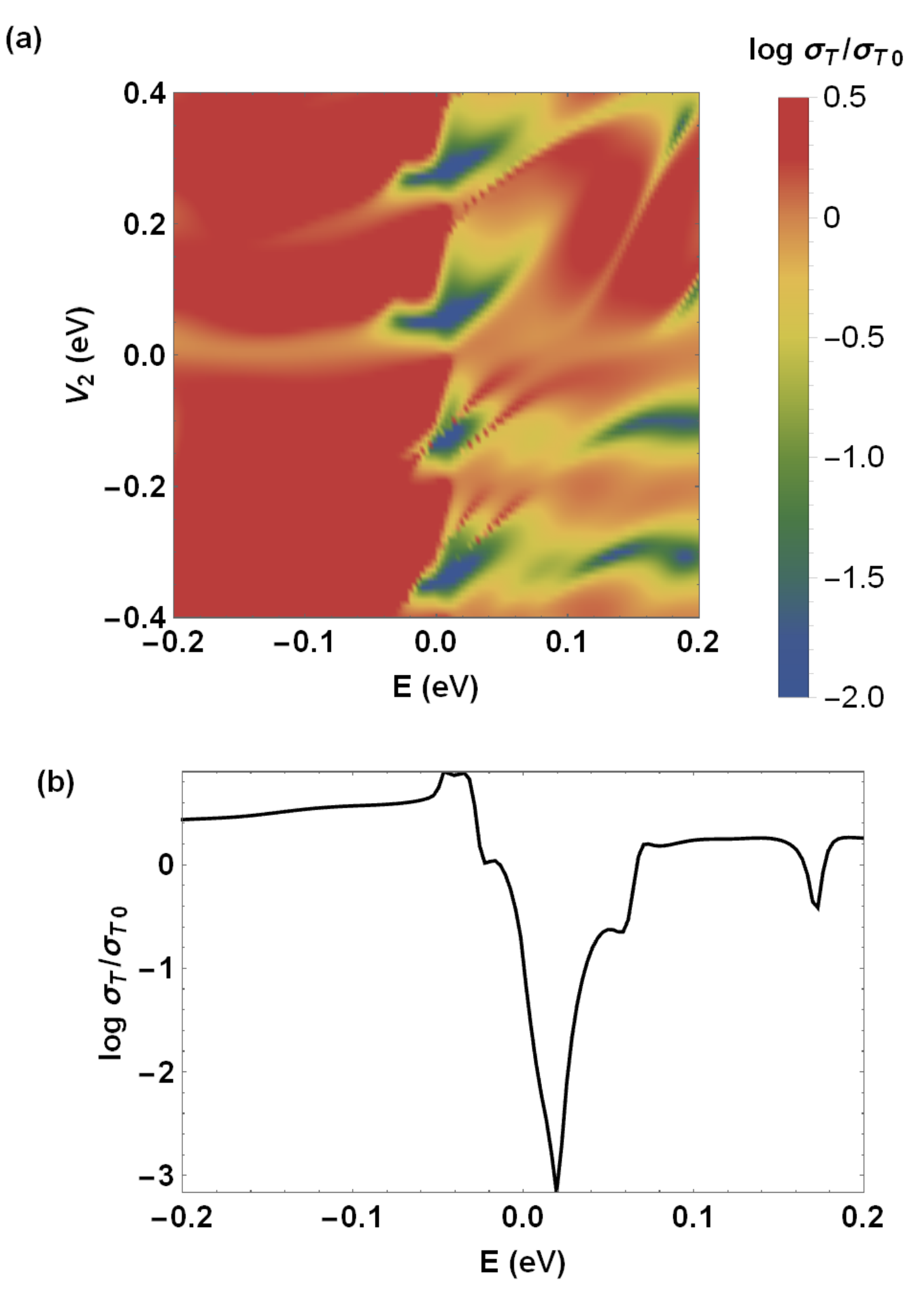}
\caption{ (a) Cloaking efficiency as a function of $V_2$ and $E$  for $R_1=10\;\text{nm}$, $V_1=-114\;\text{meV}$ and $R_2=20\;\text{nm}$. (b) Cloaking efficiency as a function of  $E$  for the same parameters of (a) and $V_2=286$ meV. }
\label{fig.5}
\end{figure}

Figure \ref{fig.5} (a)  reports  $\sigma_T/\sigma_{T0}$ as a function of the energy of the incident beam $E$ and $V_2$.  We {can see} that cloaking is more efficient at low energies, close to the Dirac point, where the changes in transport cross-section are more pronounced. If now $V_2$ is set to minimize $\sigma_T/\sigma_{T0}$, Fig. \ref{fig.5} (b) shows the cloaking efficiency as a function of the energy and highlights the energy selective nature of cloaking: The original resonance was located at $E=0.02$ eV, where the cloaking is more efficient. 

\begin{figure}[h]
\centering
\includegraphics[scale=0.5]{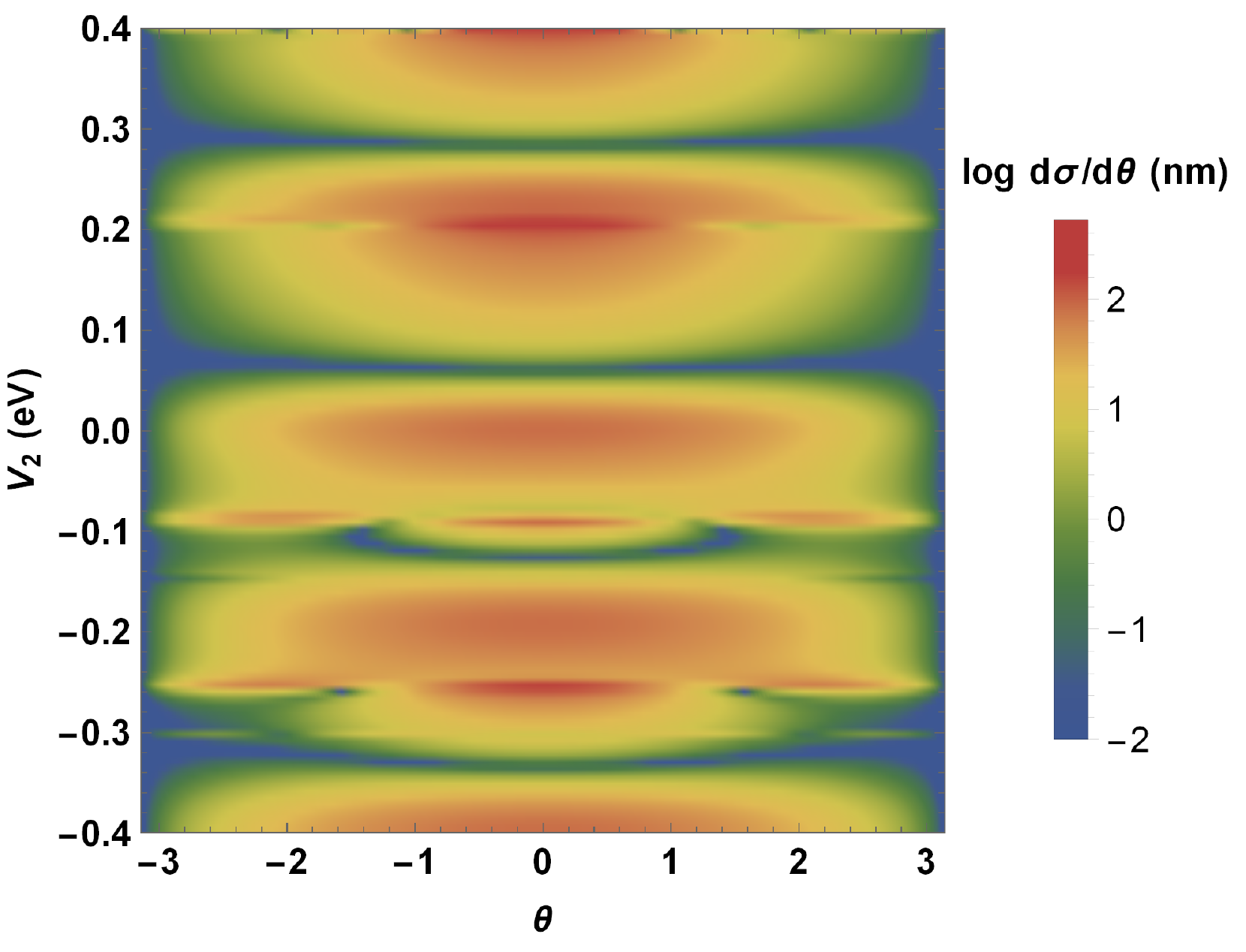}
\caption{Differential cross-section as a function of $V_2$ and the scattering angle $\theta$ for  $E=0.02\;\text{eV}$, $R_1=10\;\text{nm}$, $V_1=-114\;\text{meV}$ and $R_2=20\;\text{nm}$}
\label{fig.6}
\end{figure}

To obtain information on the direction of the scattering process, we mapped the differential cross-section as a function of the scattering angle $\theta$ and $V_1$ (see Figure \ref{fig.6} ). The absence of back-scattering is very clear for any value of $V_2$. Besides that, we can see that the cloaking is isotropic and there are changes in the scattering intensity according to the scattering angle in the resonant regime. {We conclude that tuning} $V_2$ allows the control of the scattering angle, demonstrating the feasibility of directional scattering.  

\subsection{Cloaking vacancies}
We {also investigate} the cloaking mechanism for vacancies, a different source of resonant scattering. Again we keep the original vacancy with a radius of $R_1=1$ nm and include a new potential $V_2$ in a ring of radius $R_1$ and $R_2$ ($R_2>R_1$) , or equivalently, a disk of radius $R_2$. We use the potential  
to tune  cloaking, maintaining the basic idea: we change  $V_2$ in such a way that it {leads to} $\sigma\sim0$.  The two boundary conditions $\psi_{m}^{3}(R_2,\theta)=\psi_{m}^{2}(R_2,\theta)$ and $\psi_{m}^{2}(R_1,\theta)=0$ lead to the ratio  $B_{m}/A_{m}$, and consequently $\delta_m$. 

\begin{figure}[h]
\includegraphics[scale=0.5,clip]{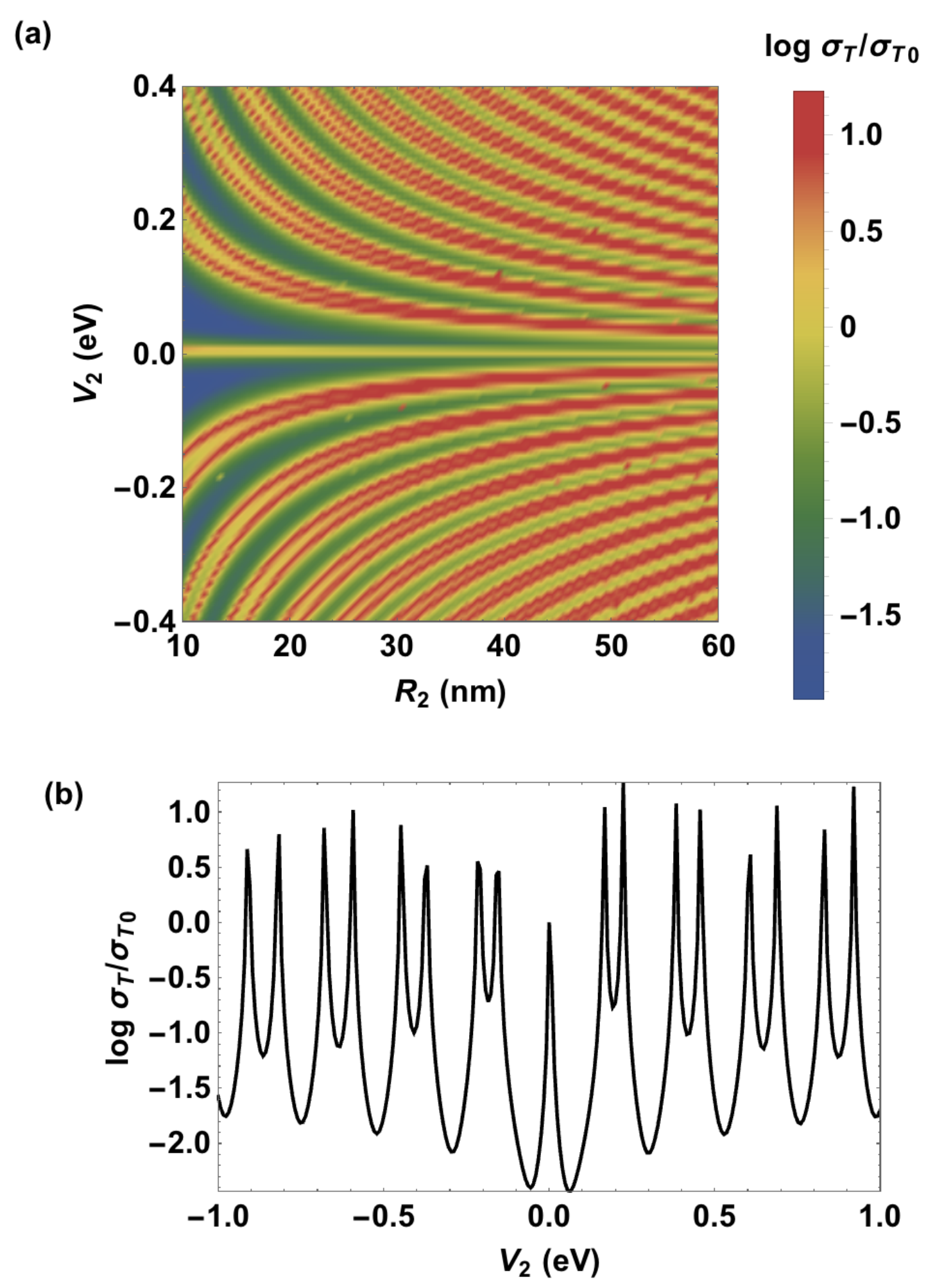}
\caption{ (a) Cloaking efficiency as a function of $V_2$ and $R_2$ for $E=2\;\text{meV}$ and a vacancy of radius $R_1=1\;\text{nm}$; (b) Cloaking efficiency for the same parameters of (a), $E=1\;\text{meV}$ and $R_2=20$ nm.}
\label{fig.7}
\end{figure}

The density plot of the cloaking efficiency as a function of $V_2$ and $R_2$, shown in Figure \ref{fig.7}a,  reveals that the most efficient cloaking occurs for rings that are 1-5 nm thick. As discussed previously, thin cloaking layers based on back and top gates are difficult to engineer. Consequently, we focus the rest of our analysis on thicker rings, of the order of $R_2 \sim \text{10-20}\;\text{nm}$. For large values of $R_2$, there are strong oscillations of the scattering intensity, going from strong scattering to transparent states with small changes in $V_2$( see Figure \ref{fig.7}b). The distance between the peaks in Figure \ref{fig.7}b is set by external radius $R_2$. {Remarkably, strongly scattering states closely coupled to invisible ones also exist in photonics systems, where the scattering cross-section of multi-layered plasmonic covers may exhibit this so-called ``comblike'' behavior~\cite{monticone2013}. Here a similar behavior in the cross-section occurs as a function of the external potential, and not as a function of the incident wavelength as in the photonics case. This result would allow one to dramatically modify electron scattering in graphene by varying an external parameter.}      

\begin{figure}[h]
\centering
\includegraphics[scale=0.55,clip]{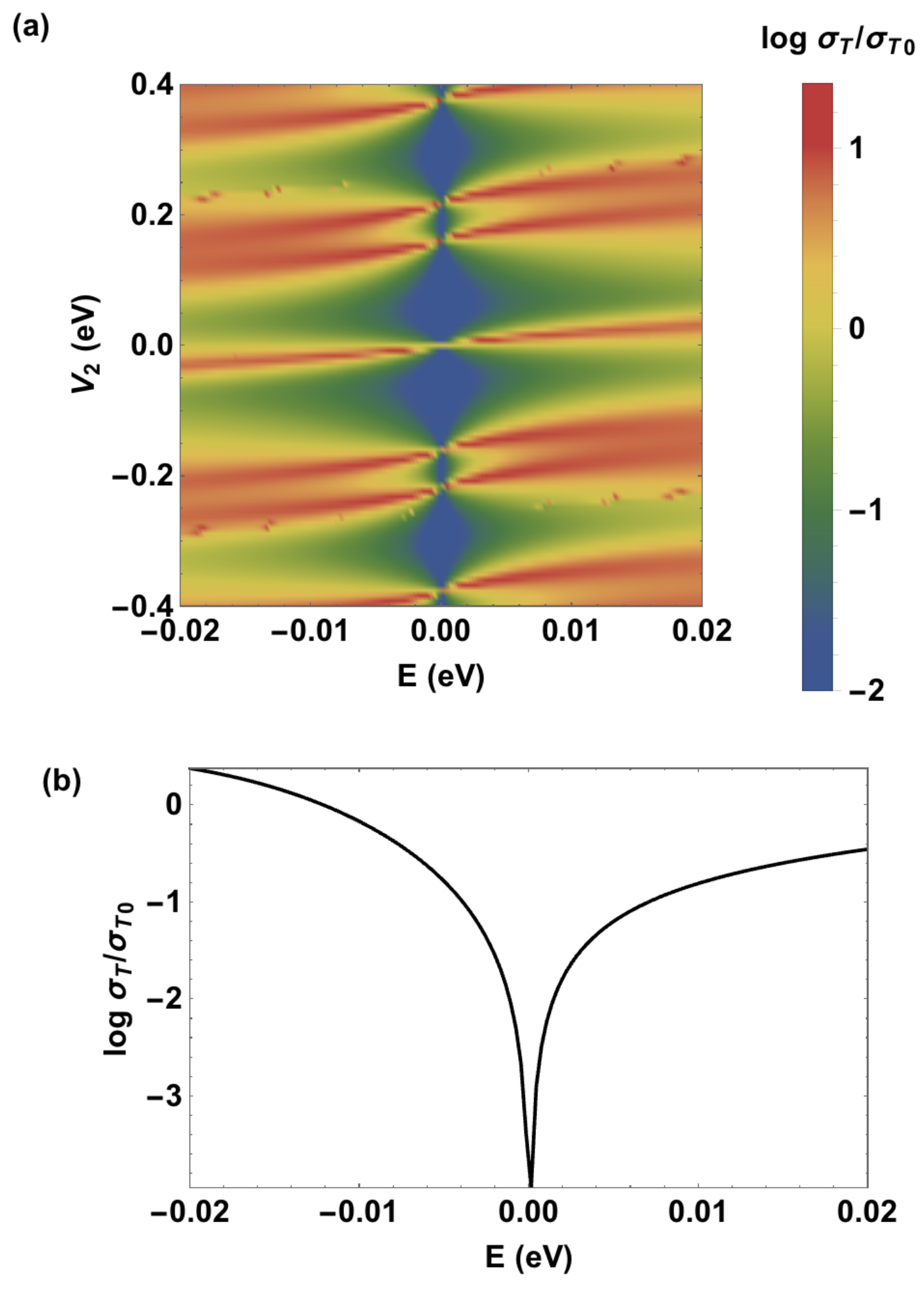}

\caption{ (a) Cloaking efficiency as a function of $V_2$ and $E$  for a vacancy of radius $R_1=10\;\text{nm}$ and $R_2=20\;\text{nm}$. (b) Cloaking efficiency as a function of  $E$  for the same parameters of (a) and $V_2=87$ meV. }
\label{fig.8}
\end{figure}

An important difference between the disk and the void {cases is} the selectivity in energy. In the first scenario, there is a reduction in the value of $\sigma_T$ of $3$ orders of magnitude at {the} specific energy where the resonant state is located, {which in turn} can be tuned by the potential $V_1$. For the vacancy, the resonance is always located at 
the Dirac point and  cloaking only occurs in the vicinity of $E=0$ (see Figure \ref{fig.8} a). For energies very close to the Dirac point,
 the cloaking is very efficient, as shown in Figure \ref{fig.8} b. However, for the energies discussed in the previous section, the transport 
cross-section is reduced to only $10\%$ of its original value.

Considering just the first partial-wave, It is possible to
 obtain a simplified expression for maximum cloaking 
($\delta_0=0$) at the Dirac point ($E=0$). The condition reads:
\begin{equation}
J_1(x)Y_0(\alpha x)-J_0(\alpha x)Y_1(x)=0,
\end{equation}
where $x=\frac{|V_2| R_2}{\hbar v_F}$ and $\alpha=\frac{R_1}{R_2}$. 
This equation is satisfied for a sequence of increasing values of $x$ for a given $\alpha<1$.  
In the limit $x\gg1$ the above equation acquires the form
\begin{equation}
 \cos[x(1-\alpha)]\approx0\,,
\end{equation}
whose solution is 
\begin{equation}
 x=\frac{(2n+1)\pi}{2(1-\alpha)}\,,
\end{equation}
with $n=0,1,2,\ldots$, 
and it is consistent with Figures \ref{fig.7} and \ref{fig.8}.

\subsection{Condition for cloaking  by an arbitrary potential}
\label{section_general}
In this subsection an integral equation for the phase-shifts is derived and a general 
condition  for the vanishing of the phase-shifts is obtained.
For a general potential $V(r)$ we seek solutions of the Dirac equation in the form
\begin{equation}
 \psi_m(r,\theta)=
\left(
\begin{array}{c}
 F(r)e^{i\theta m}\\
 iG(r)e^{i\theta (m+1)}\\
\end{array}
\right)\,.
\end{equation}
Acting with the Hamiltonian ${\cal H}_0+V$ (where ${\cal H}_0$ is the Dirac Hamiltonian) 
in the above general form of $\psi_m(r,\theta)$ we end up with two coupled
first-order differential equations:
\begin{eqnarray}
 G'+\frac{m+1}{r}G &=& \epsilon F\,,\\
 F'-\frac{m}{r}F   &=& -\epsilon G\,,
\end{eqnarray}
where $\epsilon = (E-V)/(v_F\hbar)$.
We benefit from symmetrizing the two last equations by introducing
two new functions, $G=g/\sqrt r$ and $F=f/\sqrt r$. These lead to
\begin{eqnarray}
 g'+\frac{j}{r}g &=& \epsilon f\,,\\
 f'-\frac{j}{r}f   &=& -\epsilon g\,.
\end{eqnarray}
where $j=m+1/2$.
For zero potential, we write
\begin{eqnarray}
 g_0'+\frac{j}{r}g_0 &=& \epsilon_0 f_0\,,\\
 f_0'-\frac{j}{r}f_0   &=& -\epsilon_0 g_0\,,
\end{eqnarray}
where $\epsilon_0 = E/(v_F\hbar)$.
We can now combine the two sets of equations, with and without the potential, 
by an appropriate multiplication
of the equations with the interaction by the spinor components with zero potential and vice-versa.
This procedure leads to
\begin{equation}
 \frac{d}{dr}(f_0g)=f_0f\epsilon  - gg_0\epsilon_0\,,\\
 \frac{d}{dr}(g_0f)=-g_0g\epsilon + ff_0\epsilon_0\,,
\end{equation}
which we can subtract from each other to obtain
\begin{equation}
 \frac{d}{dr}(f_0g-fg_0)=-\frac{V}{v_F\hbar}(gg_0+f_0f)\,.
\end{equation}
Integrating the preceding equation from zero to infinity and  recalling
 that $f_0$ and $g_0$ vanish at $r=0$ we obtain
\begin{equation}
 \lim_{x_0\rightarrow\infty}[f_0g-fg_0]^{x_0}=-\frac{1}{v_F\hbar}\int_0^\infty drV(r)(gg_0+f_0f)
\label{eq_integral_first}
\end{equation}
We also note that for $r\rightarrow\infty$, the scattering states are simply related to the 
asymptotic forms of the Bessel functions as
\begin{eqnarray}
f\rightarrow \sqrt{\frac{2}{k\pi}}\cos(kr-\lambda_m+\delta_m)\,,\\
g\rightarrow \sqrt{\frac{2}{k\pi}}\sin(kr-\lambda_m+\delta_m)\,.
\end{eqnarray}
Using the latter asymptotic forms, Eq. (\ref{eq_integral_first}) reads
\begin{equation}
 \frac{2}{k\pi}\sin\delta_m=-\frac{1}{v_F\hbar}\int_0^\infty drV(r)(gg_0+f_0f)\,.
\end{equation}
Now $f_0=\sqrt {r}J_m(kr)$ and $g_0=\lambda\sqrt {r}J_{m+1}(kr)$, and we obtain
\begin{equation}
 \sin\delta_m=-\frac{\pi k}{2}\frac{1}{v_F\hbar}\int_0^\infty dr\sqrt rV(r)
[fJ_m(kr) + g\lambda J_{m+1}(kr)]\,.
\label{eq_cloaking_condiction_infinity}
\end{equation}
We note that there is an alternative route for deriving the last relation, but it is
a much more complex procedure that uses the Lippamnn-Schwinger equation and the Green's
function of the Dirac Hamiltonian.
For a potential of finite range $r=R$, we have
\begin{equation}
 \sin\delta_m=-\frac{\pi k}{2}\frac{1}{v_F\hbar}\int_0^R dr\sqrt rV(r)
[fJ_m(kr) + g\lambda J_{m+1}(kr)]\,.
\end{equation}
From this equation we extract the general condition that {is} $\sin\delta_m=0$ ($\cos\delta_m=1$):
\begin{equation}
 0=\int_0^R dr\sqrt rV(r)
[fJ_m(kr) + g\lambda J_{m+1}(kr)]\equiv\tilde K_m\,,
\label{eq_cloaking_condiction}
\end{equation}
which is the central result of this subsection, since it corresponds to the general
{invisibility} condition for an arbitrary potential $V(r)$. Let us consider a simple application
of the above result taking {into account} a potential that is constant up to
$r=R$. Then Eq. (\ref{eq_cloaking_condiction}) leads to
\begin{equation}
\int_0^R drr
[J_m(pr)J_m(kr) +  J_{m+1}(pr)J_{m+1}(kr)]=0\,,
\end{equation}
where $p=(E-V)/(v_F\hbar)$ (we assumed $E>0$ and $E>V$).
Performing the integral, we obtain the condition
\begin{equation}
 0=pJ_{m-1}(pR)J_m(kR)-kJ_{m-1}(kp)J_m(pr)\,,
\end{equation}
which is the same we obtain from the phase-shift analysis.

 In the {above} form,
the result   (\ref{eq_cloaking_condiction}) is both a condition for the vanishing
of the phase-shift caused by a potential and a cloaking condition if the potential
can be separated into two (or more) different contributions that can be tuned separately, as
discussed in the previous examples. Finally it should be appreciated that condition
(\ref{eq_cloaking_condiction}) depends only on the value of the wave function in the 
region where the potential is finite.
In the next subsection we discuss an intriguing problem, the ability of a spatially short-range 
potential to cloak a resonant scatterer.

\subsection{Cloaking  a void with a $\delta-$function potential}
\label{section_application}
Let us consider 
the cloaking of a void by a $\delta-$function potential of the form $V(r)=V_0\delta(r-R_0)$;
we also define $v_0=V_0/(v_F\hbar)$.
The fact that the Dirac equation is a first-order differential equation and the 
$\delta-$potential is rather singular implies that the boundary condition cannot be the continuity of the 
spinor at the location of the potential.  The radial part of the Dirac equation can be written
in matrix form as
\begin{equation}
\frac{d{\cal S}}{dr}=-i\epsilon\sigma_y{\cal S} +i\sigma_yv{\cal S}-\frac{j}{r}\sigma_z{\cal S}
\equiv\hat G(r){\cal S}\,, 
\end{equation}
where ${\cal S}^T=(f,g)$ and $v=V/(v_F\hbar)$. Formally, the latter equation can be integrated as
\begin{equation}
{\cal S} = {\cal T}_re^{\int_{r_0}^rdr'\hat G(r') }{\cal S}(r_0)\,,
\end{equation}
where ${\cal T}_r$ is the radial position-ordering operator, defined
as 
\begin{eqnarray}
 {\cal T}_r[\hat G(r_1)\hat G(r_2)]&=&\theta(r_1-r_2)\hat G(r_1)\hat G(r_2)
\nonumber\\
&+&
\theta(r_2-r_1)\hat G(r_2)\hat G(r_1)\,.
\end{eqnarray}
Performing the integral, which is now a well defined quantity,
the spinor right after the position of the $\delta-$function
reads
\begin{equation}
 {\cal S}(R_0+\eta)={\cal T}_re^{\int_{R_0-\eta}^{R_0+\eta}dr'\hat G(r') }{\cal S}(R_0-\eta)
=e^{i\sigma_yv_0}{\cal S}(R_0-\eta)\,,
\label{eq_BC}
\end{equation}
 and $\eta=0^+$. Eq. (\ref{eq_BC}) embodies the boundary 
condition obeyed by the spinor at a $\delta-$function potential. Using Lagrange-Silvester
formula, we can compute the value of the exponential as follows: for a function $f({\cal M})$ 
[one should not confuse the spinor component $f$ with $f({\cal M})$]
of a
$2\times2$ matrix $\cal{M}$ we have that
\begin{equation}
 f({\cal M})=f(m_1)\frac{m_2-{\cal M}}{m_2-m_1}+f(m_2)\frac{m_1-{\cal M}}{m_1-m_2}\,,
\end{equation}
{where} $m_{1/2}$ are the two eigenvalues of the matrix $\cal{M}$. Applying this result to our case
we obtain
\begin{eqnarray}
 f(i\sigma_yv_0)=\cos v_0+i\sigma_y\sin v_0\,.
\end{eqnarray}
Thus, the boundary condition obeyed by the spinor reads
\begin{equation}
 {\cal S}(R_0+\eta)=(\cos v_0+i\sigma_y\sin v_0){\cal S}(R_0-\eta)\,.
\label{eq_boundary_condition}
\end{equation}
We note that in this case the spinor is not continuous, as we had anticipated,
 and therefore we need a prescription for integrating
the $\delta-$function in Eq. (\ref{eq_cloaking_condiction_infinity}). 
To that end, we have to assume that both
spinors, that to the left and that to the right of the potential, contribute half to the 
integration. Furthermore, we should not forget that the spinors at the right and at the left  
of the $\delta-$function are connected by Eq. (\ref{eq_boundary_condition}).
If this is done, then the previous method yields the 
same result as the phase-shift analysis. We should stress that for this problem
the ``free'' solutions are those referring to the scattering states in the presence 
of the void. Therefore, $J_m(kr)$ and $J_{m+1}(kr)$ in Eq. (\ref{eq_cloaking_condiction_infinity})
should be replaced by the spinor components of the scattering states in the presence
of the resonant scatterer alone, that is, by:
\begin{eqnarray}
J_{m}(kr)\rightarrow\cos\delta_m^0[J_{m}(kr)+\tan\delta_m^0Y_{m}(kr)]\,, 
\end{eqnarray}
where $\tan\delta_m^0$ is defined by Eq. (\ref{eq.12}).
In the presence of the $\delta-$function potential, 
the spinor to the right of the potential has to have the form
\begin{eqnarray}
 \psi_m(r,\theta)&=&\cos\delta_m\left[\left(
\begin{array}{c}
 J_m(kr)e^{i\theta m}\\
 iJ_{m+1}(kr)e^{i\theta (m+1)}\\
\end{array}
\right)\right.\nonumber\\
&-&
\left.
\tan\delta_m\left(
\begin{array}{c}
 Y_m(kr)e^{i\theta m}\\
 iY_{m+1}(kr)e^{i\theta (m+1)}\\
\end{array}
\right)\right]\,.
\label{Eq_spinor_right}
\end{eqnarray}
At low energies, we find the following
conditions that make the first two first phase-shifts {vanish}:
\begin{equation}
 \left\{
  \begin{array}{cc}
   \cos v_0- R_0k[1+2\ln(R/R_0)]\sin v_0=0\,,& l=0\\
  -R^2k^2\cos v_0+\frac{1}{4} R_0 (3 R^2 - 2 R_0^2)k^3 \sin v_0=0 \,,& l=1
  \end{array}
 \right.\,.
\end{equation}
Note that  for $v_0\approx\pi/2$
the cloaking condition is purely geometric: for $\delta_0=0$, we have the condition $R_0=\sqrt e R$,
whereas for $\delta_1=0$ we obtain the relation 
$R_0=\sqrt{3/2}R$. The two numbers do not differ much from each other.
Therefore, we can almost cancel the two phase shifts at the same time.
 
\begin{figure}
 \includegraphics[height=6cm]{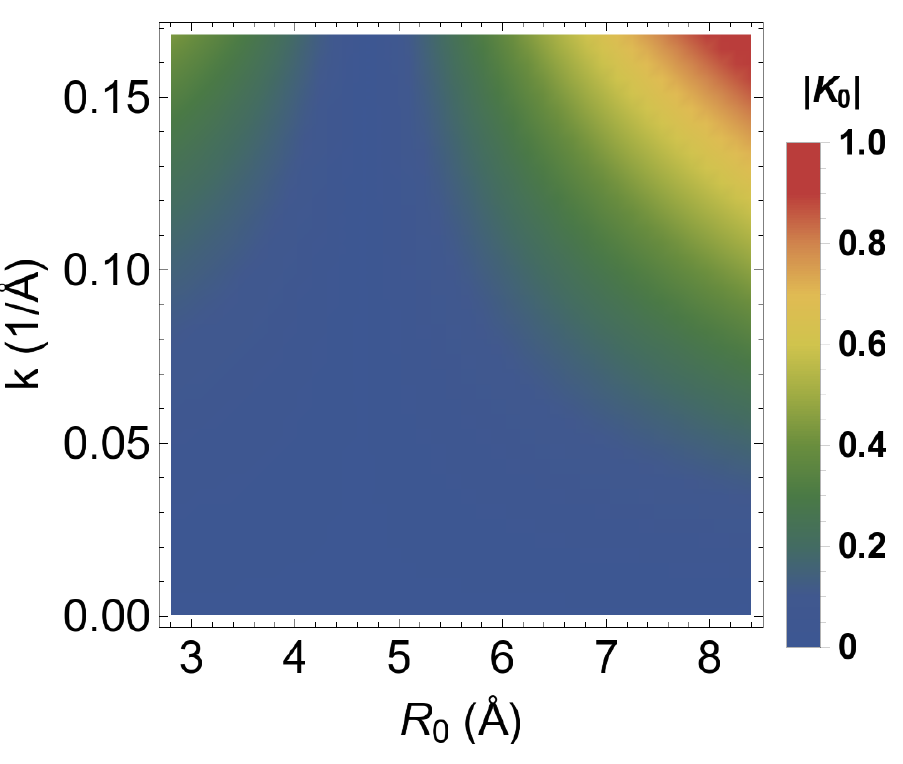}
 \includegraphics[height=6cm]{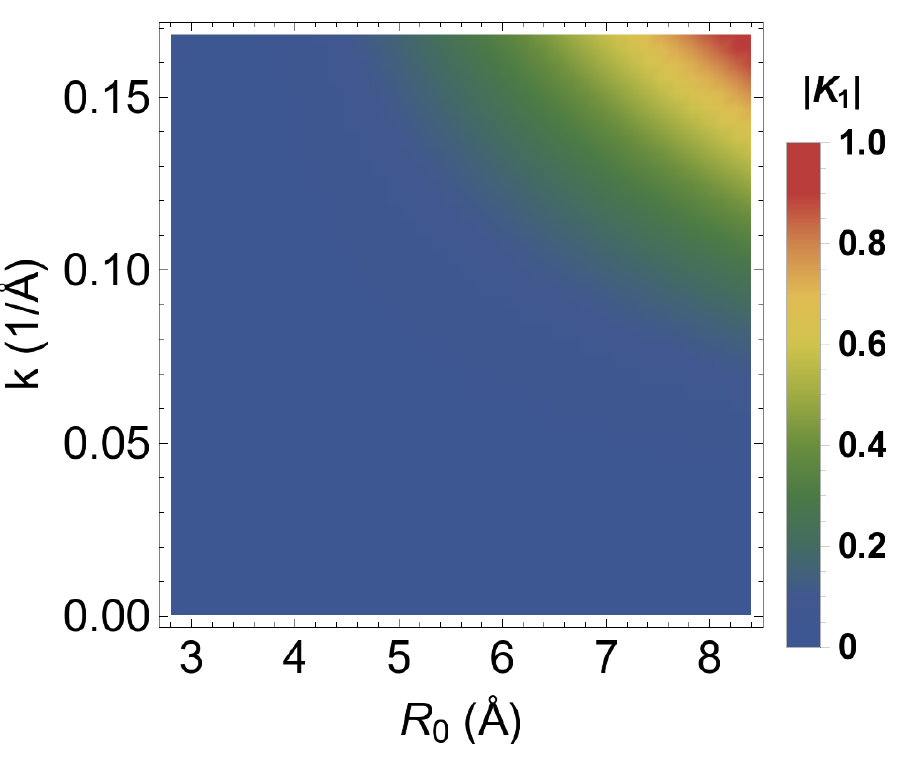}
  \caption{Density plot of $\vert\tilde  K_m\vert $, with $R=2.8$ \AA,\, and 
$v_0=\pi/2$, as function of $k$ and $R_0$. Top panel: $\vert\tilde K_0\vert$; bottom panel: 
$\vert\tilde K_1\vert$.
}
\label{fig_Km_E_R0}
\end{figure}

\begin{figure}
 \includegraphics[height=6cm]{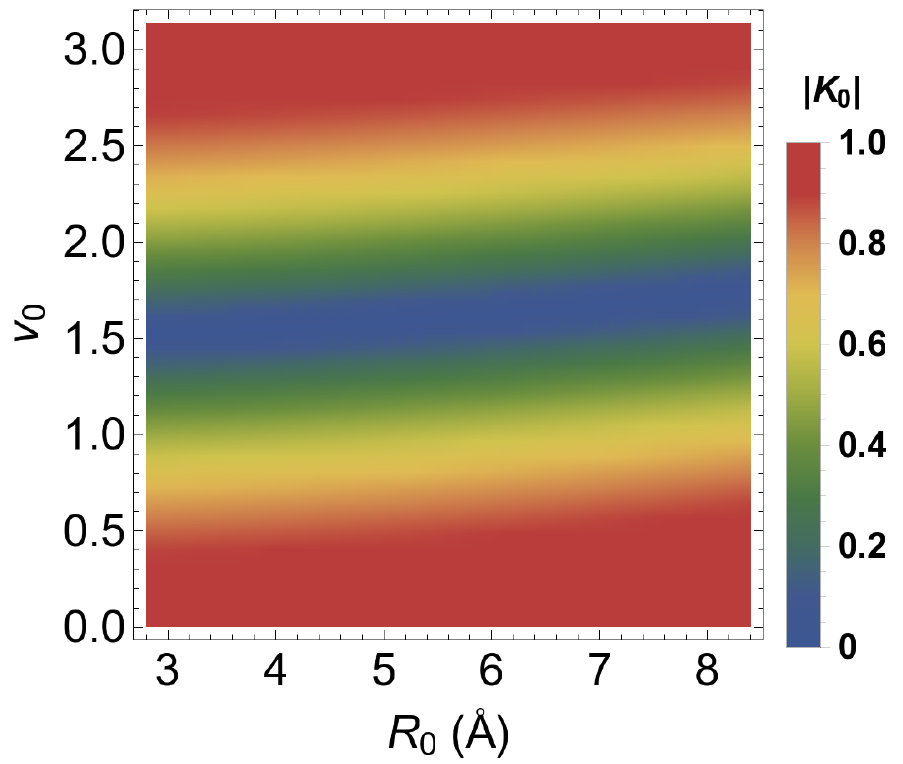}
 \includegraphics[height=6cm]{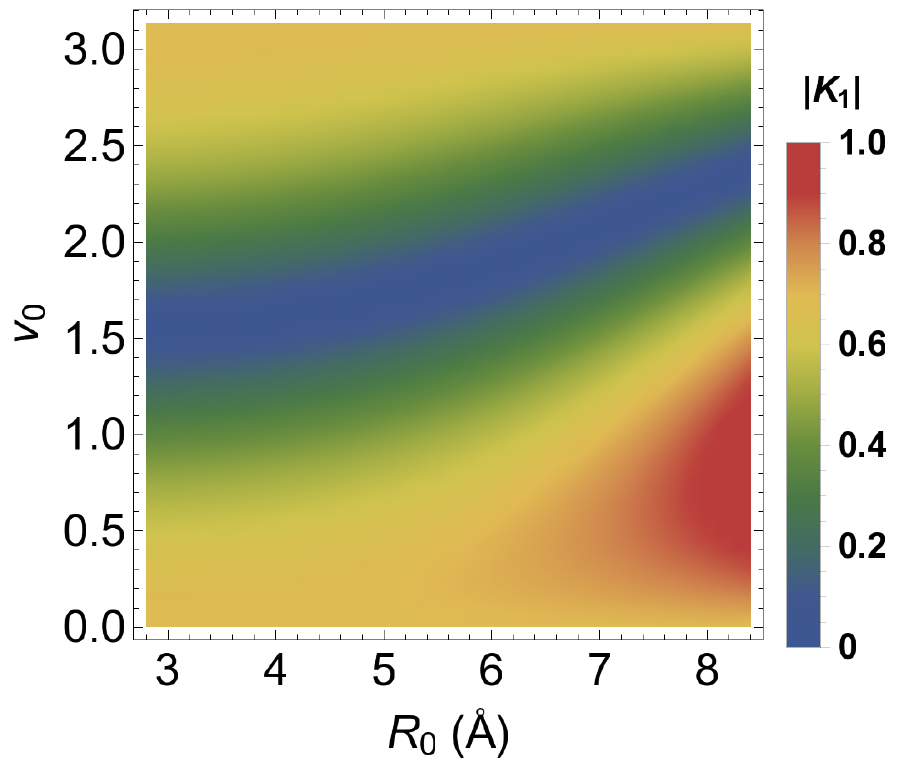}
  \caption{Density plot of $\vert\tilde K_m\vert$, with $R=2.8$ \AA,\, and 
$k=0.034$ 1/\AA\, (corresponding to $V_g=50$ V in a graphene FET on
SiO$_2$), as function of $v_0$ and $R_0$. Top panel: $\vert\tilde K_0\vert$; bottom panel:
 $\vert\tilde K_1\vert$.}
\label{fig_Km_V0_R0}
\end{figure}

In Figs. \ref{fig_Km_E_R0} and \ref{fig_Km_V0_R0} we depict  density plots of the functions
$\vert \tilde K_0\vert $ and $\vert \tilde K_1\vert$. The regions where $\vert \tilde K_m\vert=0$ 
correspond to the vanishing of
 $\sin\delta_m$ and therefore to a cloaking effect on the void by the $\delta-$function potential.
In Fig. \ref{fig_Km_E_R0} a density plot is depicted as function of $R_0$ and $k$. The vertical 
dark blue regions with $\vert \tilde K_m=0\vert $ corresponds to the $R_0=\sqrt e R$ and $R_0=\sqrt{3/2}R$
for $\vert \tilde K_0\vert $ and $\vert\tilde K_1\vert$, respectively. 
Interesting enough we see that these
geometric conditions for cloaking hold as long as $kR\lesssim1$. 
In Fig. \ref{fig_Km_V0_R0} a density plot is depicted as function of $R_0$ and $v_0$.
As in the previous figure we see dark blue regions where $\vert \tilde K_m\vert=0$ around
$v_0\approx\pi/2$.


%
\section{Tight-binding calculations\label{TB}}
To confirm our predictions and to test the robustness of the cloaking against disorder, we also model our system with a tight-binding Hamiltonian, and use the kernel polynomial method (KPM)~\cite{KPM} to calculate the local density of states of a graphene sheet~\cite{garcia2014}. This real-space approach allows us to introduce gated regions with arbitrary geometries. 
Thus, we consider the possibility of having a gate potential without perfect radial symmetry by introducing roughness to the ring potential.  The Hamiltonian reads
\be 
\mathcal{H}=-t\!\sum_{\medio{i,j}} e^{i\phi_{ij}}c^\dagger_i c_j +V_1\sum_{i \in R_1} c^\dagger_i c_i+V_2\sum_{i \in R_2} c^\dagger_i c_i
\ee
where $c_i$  is  the annihilation operator of electrons on site $i$ where $t\approx$ 2.8 eV is
the hopping energy between nearest neighbors (NN) sites in a honeycomb lattice.  $V_1$ is the radial potential  and $V_2$ is the potential of the ring.

\begin{figure}[h]
\centering
\includegraphics[width=0.80\columnwidth]{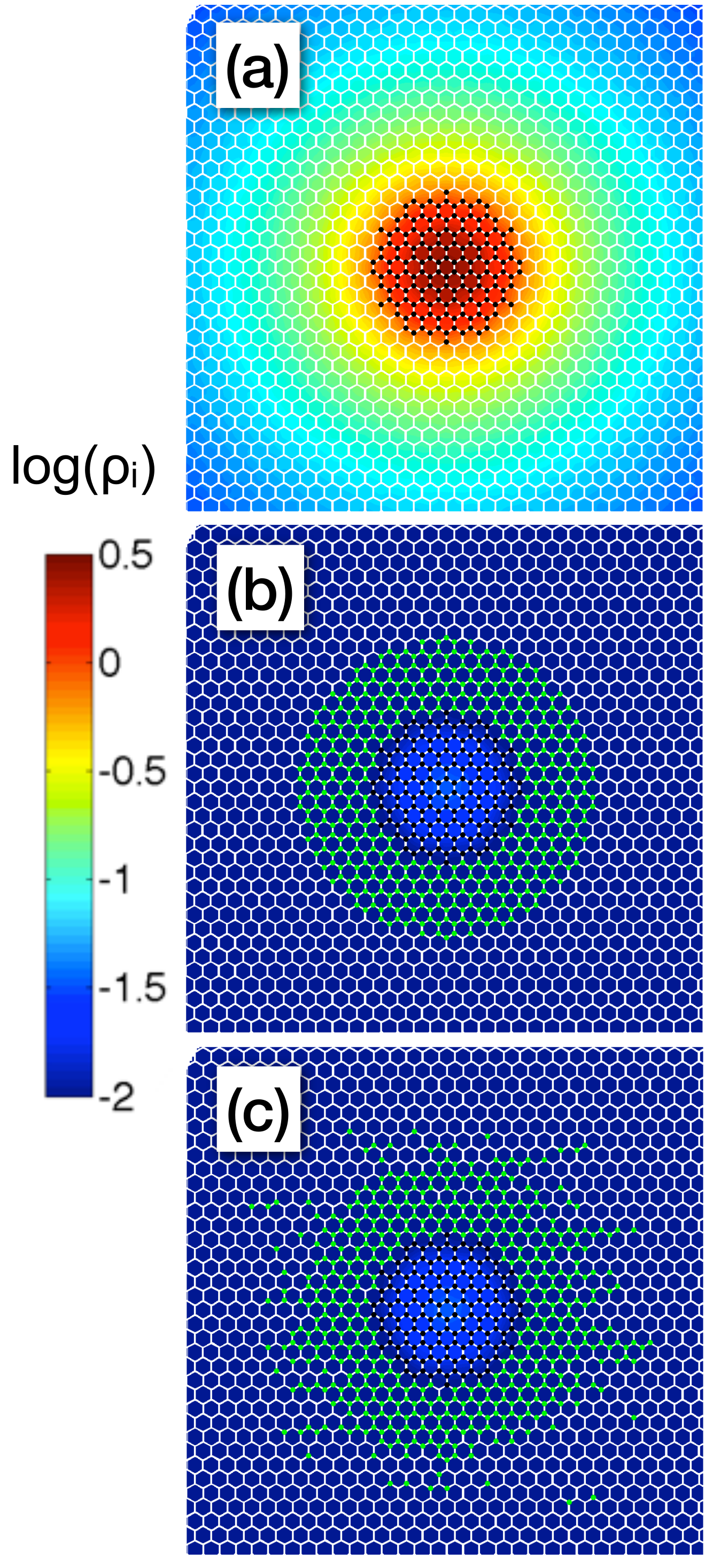}
\caption{Logarithm of the local density of states for 
 $E=0.00\;\text{eV}$. Panel (a): $V_1=0.45 t$ $V_2=0$. Panels (b) and (c) $V_1=0.45 t$ $V_2=0.23t$. 
The sites with $V_i=V_1$ are  black while the sites with $V_i=V_2$ are  green.}
\label{fig.11}
\end{figure} 

For a tight-binding hamiltonian with N=$2\times600\times 600$ sites and periodic boundary conditions, we follow the same scheme that was discussed in the previous sections: we consider a disk of radius $R_1=8a$ where $a$ is the lattice constant of graphene and tune $V_1$ to obtain a quasi-bound state inside the disk, as illustrated in Figure \ref{fig.11} for $V_1= 0.45t$, which is equivalent to the resonant scattering regime. We then vary $V_2$ inside the ring with radius $R_1=8a$ and  $R_1=16a $ to maximize the cloaking. In  Figure \ref{fig.11} b,  the quasi-bound state disappears, highlighting  the {cloaking efficiency} (even in the presence of disorder, as we shall see in {in the following}). Finally, we introduce roughness in the external side of the ring and compare the local density of states with (Figure \ref{fig.11} c) and without disorder (Figure \ref{fig.11} b). The cloaking is highly efficient in both situations,
 demonstrating the robustness of our results against disorder, which is intrinsic to the fabrication
 process. The same procedure can be repeated in the case of the vacancy and the results 
 were found to be similar. 
 
 \section{Conclusions}
In conclusion, we have envisaged novel, 
alternative electronic cloaking mechanisms in graphene. 
These mechanisms are based on drastically reducing the 
scattering cross-section of resonant scatterers in graphene,
 which can be either an engineered resonant state produced by a
 top gate or a vacancy.  Using a partial-wave expansion of the 
electronic wave-functions in the continuum approximation, described by
 the Dirac equation, we  demonstrate that electronic cloaking is very 
efficient; the reduction in the scattering cross-section can be as impressive
 as 3 orders of magnitude. In the particular case of the resonant state, we show
 that cloaking is strongly facilitated by the fact that the first phase shifts, 
which give the dominant contribution to electron scattering, are identical for graphene.
 Hence one just need to tune one of the scattering potentials to reduce significantly 
the scattering 
cross-section, this in contrast to other electronic cloaking schemes in graphene proposed so 
far~\cite{liao12}. We also show that sharp transitions between invisible and strongly 
scattering states occur for some values of the potential. We suggest that this result, 
which can be viewed as an electronic analogue of optical comblike states in coated 
scatterers~\cite{monticone2013}, could be explored in applications aiming at tuning 
the electron flow in graphene. {Using a symmetrized version of the massless Dirac equation, we derive a general
condition for the cloaking of a scatterer by a potential with radial symmetry.} 
By means of tight-binding calculations, we show that
 our findings are robust against the presence of disorder.  Altogether, our results 
reveal that electron cloaks are not only more easily achieved in graphene, but also 
that they could be applied to selectively tune the direction of electron flow in graphene
 by applying external gate potentials. As a result, we hope that could serve as the basis
 for designing novel, {disruptive} electronic devices.     

\section*{Acknowledgements}
NMRP acknowledges support from EC under Graphene Flagship (Contract No.
CNECT-ICT-604391), the hospitality of the Instituto de F\'{\i}sica of the UFRJ, 
and stimulating discussions with Bruno Amorim on the Lippamnn-Schwinger equation for
Dirac electrons. TGR thanks the Brazilian agencies CNPq and FAPERJ and Brazil Science without Borders program for financial support. FAP acknowledges CAPES (Grant No. BEX 1497/14-6) and CNPq (Grant No. 303286/2013-0) for financial support.

\end{document}